\documentstyle[aps,epsf,eqsecnum,feynmp,graphics,amsfonts, amssymb, latexsym]{revtex}
\begin{document}
\title{Exact Solutions of the Time Dependent Schr\"odinger Equation in One Space
Dimension.}
\author{B.~Hamprecht}
\address{Institut für Theoretische Physik, Freie Universität Berlin,\\
Arnimallee 14, D-14195 Berlin, Germany\\
{\scriptsize e-mails: bodo.hamprecht@physik.fu-berlin.de}
\date{May 16th, 1997}}

\maketitle

\abstract{A closed expression for the harmonic oscillator wave function after the 
passage of a linear signal with arbitrary time dependence is derived. Transition 
probabilities are simple to express in terms of Laguerre polynomials. 
Spontaneous transitions are neglected. The exact result is of some interest for the
physics of short laser pulses, since it may serve as an accuracy test for numerical
methods.}

\section{Construction of the Propagator}
We consider the Schr\"odinger Equation:
\begin{equation}
\label{TDSEQ}
-\frac{\hbar^2}{2m}\frac{\partial^2\Psi(x,t)}{\partial x^2} +
\left (\frac{m\omega^2}{2} x^2 + xj(t)\right )\Psi (x,t)= 
i\hbar\frac{\partial\Psi (x,t)}{\partial t}
\end{equation}
The driving force $j(t)$ is supposed to be real and of finite duration. The 
propagator of the above equation is well known ~\cite{BK}.
\begin{equation}
\label{PROP}
\Psi (x,t,y) = \frac{ \alpha}{\sqrt{2\pi i\sin\omega t}} \exp\left\{ \frac{i}
{\sin\omega t}\left [ \alpha^2 \left ( \frac{x^2 + y^2}{2}\cos\omega t
 - xy \right )
+\frac{xx_0 + yy_0}{\hbar}+ \frac{\chi}{2\alpha^2\hbar^2}\right ] \right \} 
\end{equation}
\noindent
where:
\begin{equation}
x_0 = -G,\qquad y_0 = G\cos\omega t -F\sin\omega t , \qquad chi = G^2 \cos\omega t-
(FG+2H)\sin\omega t , \qquad \alpha = \sqrt{\frac{m\omega}{\hbar}}
\end{equation}
\begin{eqnarray}
G(t) & = & \int\limits_{0}^{t} dt' j(t')\sin\omega t' \, , \qquad
F(t) = \int\limits_{0}^{t} dt' j(t' )\cos\omega t' \nonumber\\
H(t) & = & \frac{1}{2}\int\limits_{0}^{t} dt' j(t')
\int\limits_{0}^{t'} dt'' j(t'' )\sin\omega(t'' - t' )\nonumber
\end{eqnarray}
This path integral result describes the propagation of a Gaussian wave packet, 
which starts from $\delta$-function shape at time $t=0$. More easily it may be
derived by solving the time dependent Schr\"odinger equation (\ref{TDSEQ}).\\
\\
We notice that the wave packet
\begin{equation}
\label{SOL}
\Psi(x,t)=\exp\left\{ -A(t)\frac{\alpha^2 x^2}{2} + i B(t) \alpha x-
\frac{C(t)}{2}\right\}
\end{equation}
solves equation (\ref{TDSEQ}), if the complex valued functions $A(t), B(t)$ and 
$C(t)$ are solutions 
of the following set of ordinary differential equations:
\begin{equation}
\label{ABC}
\frac{d}{dt}A = i\omega (1-A^2)\qquad \frac{d}{dt}B = -i\omega A\, B -
\frac{j(t)}{\hbar\alpha}\qquad \frac{d}{dt}C = i\omega (A+B^2)
\end{equation}
Since we want to have a $\delta$-function initially:
$$
\Psi (x,0,y):= \Psi (x,0) = \delta (x-y) = \lim\limits_{\tau\rightarrow 0}
\frac{1}{\sqrt{\pi\tau}} \exp \left ( -\frac{(x-y)^2}{\tau} \right )
$$
$A,B$ and $C$ have to be solutions of (\ref{ABC}) with a singularity at $t=0:$
\begin{eqnarray}
\label{RABC}
A(t) & = & -i\cot\omega t\nonumber \\
B(t) & = & -\frac{\frac{G(t)}{\hbar}+\alpha^2 y}{\alpha\sin\omega t}\nonumber \\
C(t) & = & \log (2\pi i\sin\omega t) - i\alpha^2 y^2 \cot\omega t - i
\left ( 2y + \frac{G}{\alpha^2\hbar}\right )\frac{G\cot\omega t - F}{\hbar} +
\frac{2i}{\alpha^2\hbar^2}H
\end{eqnarray}
Substituting eqs. (\ref{RABC}) in equ. (\ref{SOL}) yields the Propagator (\ref{PROP}).
\section{Transition Probabilities}
Transition probability amplitudes between energy eigenstates of the harmonic 
oscillator are given by:
\begin{eqnarray}
\label{COEFF}
a_{m,n} & = & \int\limits_{-\infty}^{\infty}\psi_m^* (x,t)\Psi (x,t,y)\psi_n
(y,0)dx\, dy \nonumber\\
        & = & \alpha\int\limits_{-\infty}^{\infty}\frac
{H_m (\alpha x)\Psi (x,y,t)H_n(\alpha y)}
{\sqrt{2^{n+m}m!n!\pi}}\exp\left\{ -\frac{x^2 +y^2}{2} \alpha^2 +i\hbar\omega t
\left ( m+\frac{1}{2}\right )\right\}dx\, dy\nonumber \\
\end{eqnarray}
This integral is evaluated making use of the generating function
$\exp (2zx - z^2 )$ for the Hermite polynomials $H_n (x)$ (cf. Appendix), 
yielding for $m\leq n$:
\begin{equation}
\label{MR}
a_{n,m} = \sqrt{\frac{m!}{n!}}\exp\left\{ -\frac{|r|^2}{2}-\frac{iH}
{\alpha^2\hbar^2}\right\} (-ir^{\ast} )^{n-m} L_m^{(n-m)} (|r|^2 )
\end{equation}
where $L_m^{(n-m)} (x)$ is a generalised Laguerre polynomial\\
\\
and 
$$ r=\frac{F-iG}{\sqrt{2} \alpha\hbar} = \frac{1}{\sqrt{2} \alpha\hbar}
\int\limits_0^t dt' j(t')e^{i\omega t'}
$$
Equation (\cite{MR}) is our main result.\\
For transitions from the ground state, i.e. $m=0$, equation (8) reduces to the
simple expression:
\begin{eqnarray}
a_n & = & \frac{1}{\sqrt{n!}} \exp\left\{ -\frac{|r|^2 + iH}{2}\right\}
(-ir^{\ast} )^n\nonumber \\
|a_n|^2 & = & \frac{R^n}{n!} e^{-R} \qquad \mbox{where} \qquad R=|r|^2
\end{eqnarray}
\section{Exitation of Wave Packets}
If the pulse $j(t)$ is applied to the ground state of the harmonic oscillator, it
will produce a wave packet, which is easily determined, applying the propagator
to the wave function of the ground state:
\begin{equation}
\Psi (x,t) = \sqrt[4]{\frac{\alpha^2}{\pi}}\int\limits_{-\infty}^{\infty}
\Psi(x,t,y)\exp\left ( -\frac{\alpha^2 y^2}{2}\right ) dy
\end{equation}
The integral evaluates to:
\begin{equation}
\Psi (x,t) = \sqrt[4]{\frac{\alpha^2}{\pi}}\exp \left [ -\frac{\alpha^2}{2}
\left ( x-x_0)^2\right ) -\chi(t)\right ]
\end{equation}
\noindent
where
\begin{eqnarray}
x_0 & = & -i\frac{(F+iG)e^{-i\omega t}}{\hbar\alpha^2}\nonumber\\
\chi(t) & = & \frac{(F+iG)(F\cos\omega t + G\sin\omega t)e^{-i\omega t}+2iH}
{2\alpha^2\hbar^2}+\frac{1}{2}\;i\omega t \nonumber
\end{eqnarray}
describing an oscillating Gaussian wave packet of constant width 
$\Delta x^2 = 1/(2\alpha^2 )$.\\
The expectation values for position and momentum
turn out to be:
\begin{equation}
\langle x\rangle = \frac{G\cos\omega t - F\sin\omega t}{\alpha^2\hbar}
\qquad \langle p\rangle = -(F\cos\omega t + G\sin\omega t)
\end{equation}
\section{Appendix}
We evaluate equation (\ref{COEFF}), using the generating function for the Hermite
polynomials. If $\gamma_{m,n}$ is the coefficient of $w^m z^n$ in the Taylor
expansion of:
$$
J = \alpha \int\limits_{-\infty}^{\infty} \Psi (x,t,y)e^{2(wx+zy)\alpha -
w^2-z^2-\frac{x^2 + y^2}{2}\alpha^2} dx \,dy  \qquad \mbox{then}
$$
\begin{equation}
a_{m,n} = \sqrt{\frac{n!m!}{2^{m+n}\pi}}\exp\left\{ i\hbar\omega t
\left ( m+\frac{1}{2}\right ) \right\} \gamma_{m,n}
\end{equation}
Now:
\begin{eqnarray}
J & = & \frac{1}{\sqrt{2\pi i\sin t}}\int\limits_{-\infty}^{\infty}\exp\left\{
-X^T AX +2P^T \, X-Q\right\} \alpha^2 dX \nonumber\\
& = & \sqrt{\frac{\pi}{2i\sin t det A}}\exp\left\{ P^T A^{-1}P-Q\right\}
\end{eqnarray}
where:
\begin{eqnarray}
X & = & \alpha{x\choose y}\qquad
P={w-i\frac{G}{2\alpha\hbar\sin\omega t}
\choose z-\frac{i}{2\alpha\hbar}(F-G\cot\omega t)}\qquad
A=\left ( \begin{array}{cc}
\frac{1}{2}-\frac{i}{2}\cot\omega t & \frac{i}{2\sin\omega t}\\
\frac{i}{2\sin\omega t} & \frac{1}{2}-\frac{i}{2}\cot\omega t
\end{array}\right )\nonumber\\
Q & = & w^2 + z^2 + \frac{i}{2\alpha^2\hbar^2} (F-G\cot\omega t)
G + \frac{i}{\alpha^2\hbar^2}H \nonumber
\end{eqnarray}
We find:
\begin{equation}
det\, A = -i\frac{e^{i\omega t}}{2\sin\omega t} \qquad A^{-1} =\left (
\begin{array}{cc}
1 & e^{-i\omega t}\\
e^{-i\omega t} & 1
\end{array}\right )
\end{equation}
\begin{equation}
P^TA^{-1}P-Q=2wz \, e^{-i \omega t}-iz\frac{F-iG}{\alpha\hbar}-iw\frac{F+iG}
{\alpha\hbar}e^{-i\omega t}-\frac{F^2 + G^2 + 4iH}{4\alpha^2\hbar^2}
\end{equation}
Therefore:
$$
a_{m,n}=\sqrt{\frac{n!m!}{2^{n+m}}}\exp\left\{ -\frac{F^2 + G^2 +4iH}{4\alpha^2\hbar^2}
\right\} \sum_{k=0}^m\frac{\left( 2e^{-i\omega t}\right )^k
\left ( \frac{F+iG}{\alpha\hbar}\right )^{m-k}
\left ( \frac{F-iG}{\alpha\hbar}\right )^{n-k}(-i)^{m+n-2k}e^{ik\omega t}}
{k!(m-k)!(n-k)!}
$$
Extracting common factors from the sum and replacing $k$ by $m-k$ in the
summation, we obtain:
$$
a_{n.m}=\sqrt{n!m!}\exp\left\{ -\frac{F^2 +G^2 +4iH}{4\alpha^2\hbar^2}\right\}
\left ( -i\frac{F-iG}{\sqrt{2}\alpha\hbar}\right )^{n-m}
\sum_{k=0}^m \frac{\left ( -\frac{F^2 + G^2}{2\alpha^2\hbar^2}\right )^k}
{k!(m-k)!(n-m+k)!}
$$
The finite sum in this expression defines a Laguerre polynomial. Therefore:
\begin{equation}
a_{n.m}=\sqrt{\frac{m!}{n!}}\exp\left\{ -\frac{|r|^2}{2}-\frac{iH}{\alpha^2
\hbar^2}\right\} (-ir^{\ast} )^{n-m}L_m^{(n-m)}(|r|^2)
\end{equation}
where:
$$
r=\frac{F+iG}{\sqrt{2}\alpha\hbar}= \frac{1}{\sqrt{2}\alpha\hbar}
\int\limits_{0}^{t}dt'j(t')e^{i\omega t'}
$$
which proves equation (\ref{MR}).

\begin{thebibliography}{1}
%
\bibitem{BK}
Hagen Kleinert:
{\bf Path Integrals in Quantum Mechanics, Statistics and Polymer Physics}, \,
\, $2^{nd}$
edition, World Scientific(1995). 
%
\end{thebibliography}
\end{document}